\RequirePackage{ifpdf}
\documentclass{JINST}

\title{Fibre laser hydrophones for cosmic ray particle detection}

\author{E.~J.~Buis$^a$\thanks{Corresponding
    author.}~, E.~J.~J.~Doppenberg$^a$, R.~A.~Nieuwland$^a$ and  P.~M.~Toet$^a$\\
  \llap{$^a$}TNO Optics,\\
  Delft, the Netherlands\\
  E-mail: \email{ernst-jan.buis@tno.nl}}

\abstract{The detection of ultra high energetic cosmic neutrinos provides a unique means to search for extragalactic sources that accelerate particles to extreme energies. It allows to study the neutrino component of the GZK cut-off in the cosmic ray energy spectrum and the search for neutrinos beyond this limit. Due to low expected flux and small interaction cross-section of neutrinos with matter large experimental set-ups are needed to conduct this type of research. Acoustic detection of cosmic rays may provide a means for the detection of ultra-high energetic neutrinos. Using relative low absorption of sound in water, large experimental set-ups in the deep sea are possible that are able to detect these most rare events, but it requires highly sensitive hydrophones as the thermo-acoustic pulse originating from a particle shower in water has a typical amplitude as low as a mPa. It has been shown in characterisation measurements that the fibre optic hydrophone technology as designed and realised at TNO provides the required sensitivity. Noise measurements and pulse reconstruction have been conducted that show that the hydrophone is suited as a particle detector.}


\begin{document}

\section{Motivation}
\label{chap:intro}
With the discovery of the extraterrestrial origin of radio-active background radiation by Victor Hess in 1912, the field of cosmic ray research is now over hundred years old. In this period a large variety of detection methods have been developed to study the cosmic rays, ranging from balloon born instruments to a multi-ton instrument on board the ISS. To study the high-end of the cosmic-ray energy spectrum large volume experimental set-ups are required as the event rate drops with the cosmic ray particle energy. In particular the energy scale above 10$^{18} - 10^{19}$ eV is of special interest. It has been predicted to be the ultimate energy known as the GZK cut-off energy, named after Greisen, Zatsepin and Kuzmin, who published on the topic \cite{GKZ} in the 1960's. This cut-off energy is induced by the interaction cross-section of protons with the cosmic microwave background radiation (CMB) that increases with energy, and which reduces the absorption length of protons in the universe to galactic length scales. This energy limit has been first measured by HiRes experiment \cite{HiRes} and confirmed by the Pierre Auger experiment \cite{Auger}. Complementary, experiments are planned and being built, dedicated to the study of ultra-high energy cosmic neutrinos. The study of cosmic neutrinos of the highest energies, the ultra-high energy neutrinos, may provide information of the neutrino component of the GZK cut-off and possibly a hint of energies beyond the upper limit of the energy scale. Although the  confirmation of the GZK cut-off limit firmly predicts the existence of a similar cut-off in the neurtino component, the exact shape of the neutrino flux in the tail of the energy distribution depends on several yet unknown parameters. For instance, the measurement of the ultra high energy neutrinos would learn us on the proton and ion content of the cosmic ray flux. Furthermore deviation from the expected neutrino spectrum may rise if neutrinos are produced by an unknown exotic source that cannot probed using protons.

Cosmic neutrinos are commonly studied by measuring the \u{C}erenkov radiation, when the neutrino has converted to a muon. The \u{C}erenkov radiation is then detected using light sensitive sensors, such as photo-multiplier tubes (PMT).  An established method is to place a high number of PMTS in water or ice and to direct them towards the center of the Earth to suppress the large background coming from the 'atmospheric' muons from the upper hemisphere. Such an experimental method would then be only sensitive to Earth penetrating nuetrinos. At high energies however, above 10$^{14}$ eV, the Earth becomes opaque to neutrinos due to the fact that the interaction cross-section of neutrinos increases with energy. This means that a \u{C}erenkov based experiment is not able to study the very end of the energy spectrum of cosmic neutrinos. A means to probe the ultra-high energy spectrum of cosmic neutrinos may be offered by the acoustic detection of cosmic neutrinos as explained in the next section. Acoustic detection of neutrinos would hence be a detection method that is complementary to the \u{C}erenkov based detection method.

In this paper the acoustic detection principle for cosmic rays set-ups located in the deep sea is discussed and a sensitive hydrophone technology is introduced. We report on the design, realisation and characterisation of a prototype fibre laser hydrophone. In the next section we review the cosmic neutrino induced acoustic signals and define requirements for a hydrophone technology. Then in section \ref{sec:fibres_at_TNO} we describe the hydrophone as designed and produced at our institute. Then, in section \ref{sec:results} we report on the characterisation of the hydrophone in our lab and we conclude in section \ref{sec:conclusions}.

\section{Acoustic cosmic neutrino detection}
\label{sec:acoustics}
\subsection{Acoustic signal}
\label{sec:acoustic_signal}
The mechanism of acoustic detection of energetic particles has been investigated already in the 1960's with pioneering work carried out by Askaryan et al\cite{Askaryan57}. Measurements were carried out at proton accelerators to investigate the acoustic signals induced by the energy loss of energetic particles in water. 

\begin{figure}[ht]
 \begin{minipage}[]{0.49\textwidth}
   \begin{center}
   \includegraphics[width=0.95\textwidth]{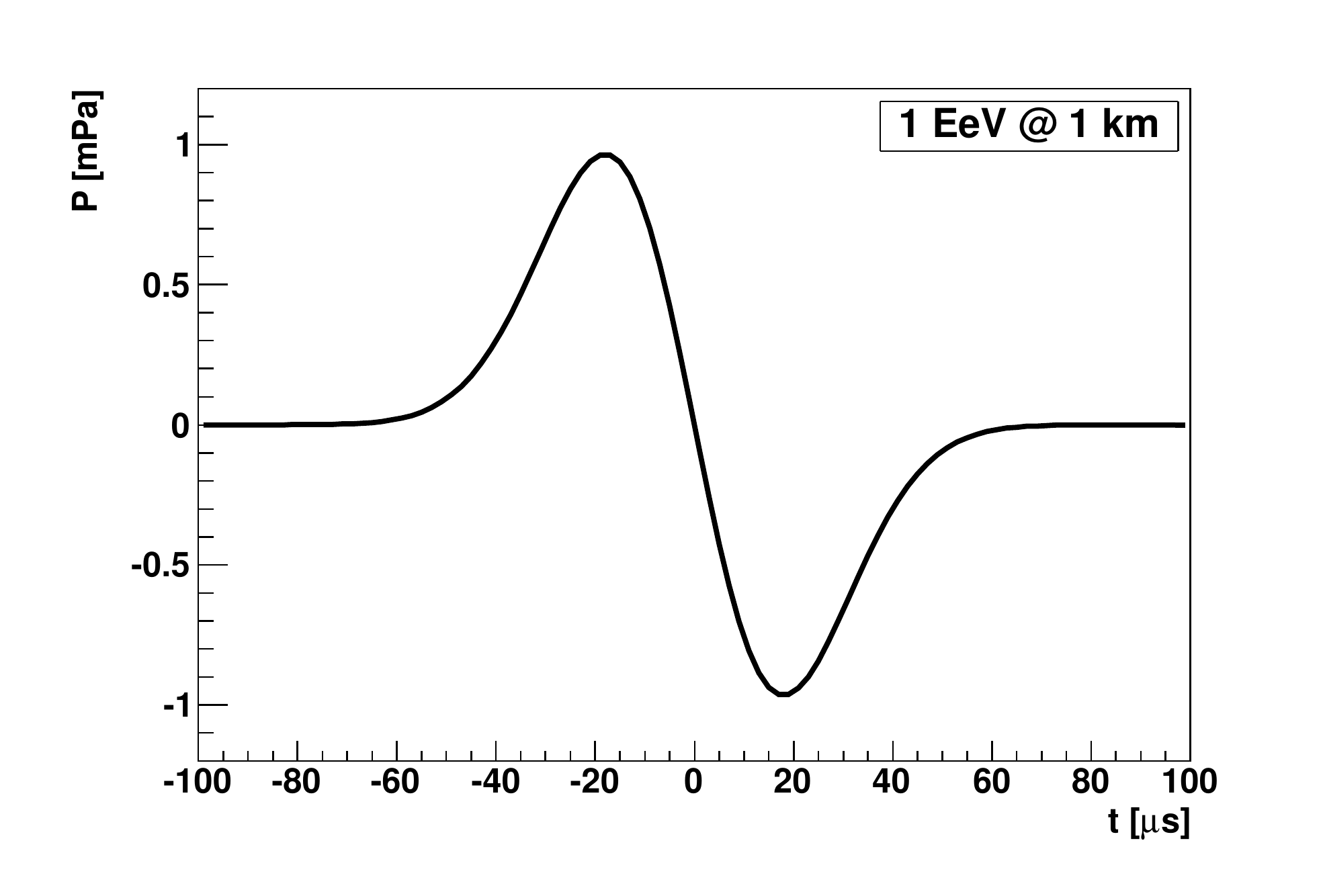}
   \end{center}
  \end{minipage}
  \begin{minipage}[]{0.51\textwidth}
   \begin{center}
  \includegraphics[width=0.95\textwidth]{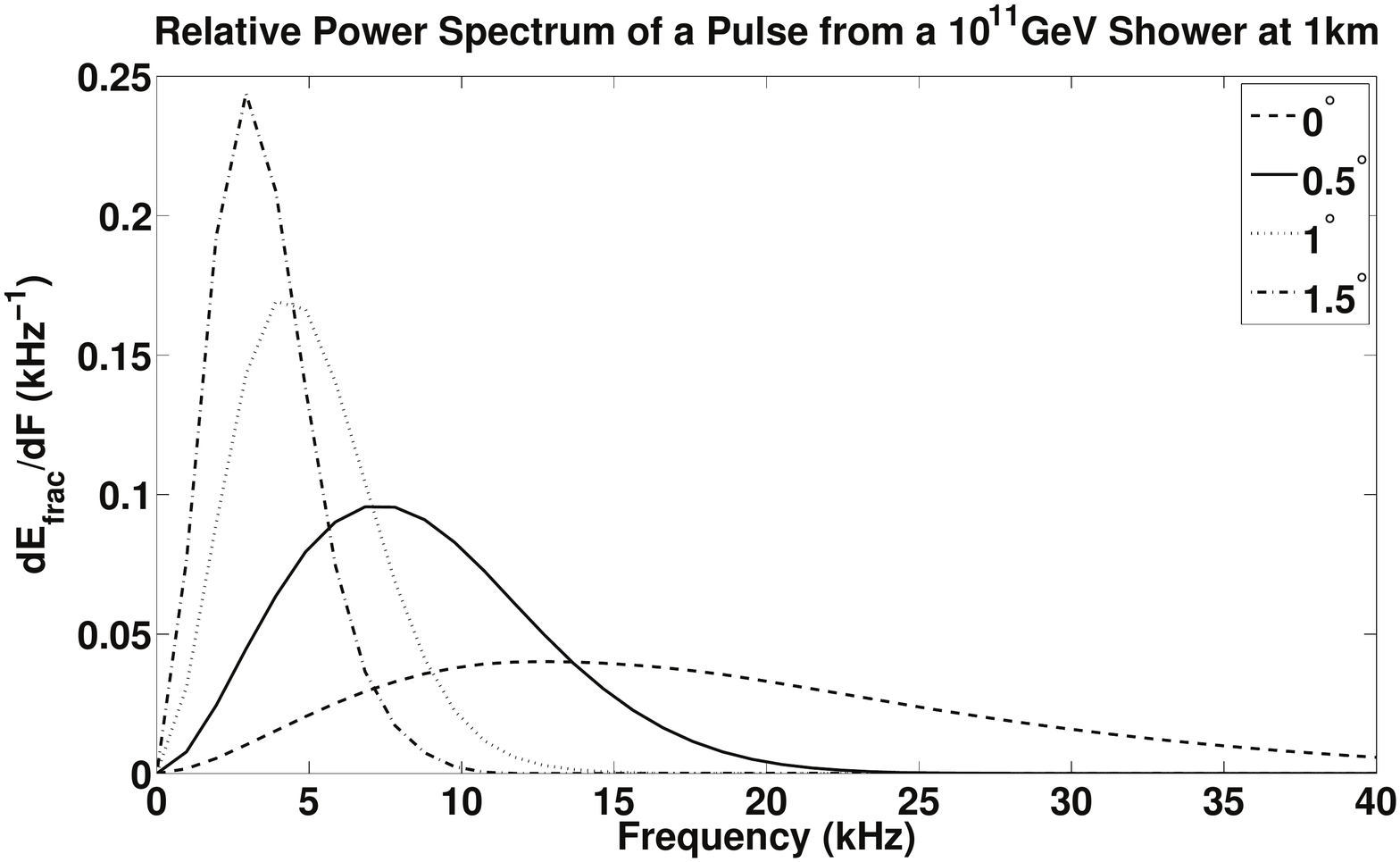}
   \end{center}
  \end{minipage}\\ \\
  \begin{minipage}[]{0.49\textwidth}
    \begin{center}
      (a)
    \end{center}
  \end{minipage}
  \hfill
  \begin{minipage}[]{0.51\textwidth}
    \begin{center}
      (b)
    \end{center}
  \end{minipage}
  \begin{center}
    \begin{minipage}[b]{\textwidth}
      \caption[]{\label{fig:signal} (a) Typical acoustic signal from a energetic neutrino induced particle shower. In (b) the relative power of a pulse as a function of the frequency. The spectrum is plotted for various angles w. r. t. to the plane perpendicular to the particle shower axis. Adapted from \cite{Lahmann11}. }
  \end{minipage}
 \end{center}
\end{figure}

In the case of neutrino detection, the particle detection is detected indirectly after a neutrino has interacted through the neutral and charged current weak interactions which induce a shower of secondary particles in water. The particle shower forms a long line acoustic source of several cm in radius and about several tens meters long. This line source gives rise to a distinct signal of which an example is shown in figure \ref{fig:signal}. The shape of the signal is a typical bipolar pulse. 

The relative power spectrum as shown in figure \ref{fig:signal} (b) shows that the expected frequency range is found between 1 and 50 kHz, with a peak power around 10 kHz. Note that the  peak in the power spectrum is closely related to the angle of the line of sight with the shower axis. The sonic field that is formed by the particle shower can be characterised as a thin, flat disk, with the shower axis forming the axis of the disk. For a typical hadronic shower, the sonic field is a disk of about 15 meter thick, while the radius of the disk can be as large as 1 km. 

The absorption length of sea water in the frequency range of interest is relatively long: At the 25 kHz the absorption length is about 1 km, while at 10 kHz it is almost 10 km \cite{Ainslie98}. These values for the absorption length indicate that indeed a kilometer sized detector would be possible. The concept of a deep sea acoustic neutrino telescope consists therefore of long strings, each equipped with several tens of hydrophones. A large number of strings would then form a fine grid of sensors with a total volume larger than a cubic kilometer. An optimized geometry of such a telescope is beyond the discussion of this paper. We mention, however, that the strings of hydrophones would consist of an oil filled hose to balance the isostatic pressure. The oil filled hose is further discussed in section \ref{sec:results}.

\subsection{Sea state noise}
\label{sec:seastatenoise}
A practical limit to the acoustic detection of cosmic neutrinos is set by the omni-present acoustic noise in the ocean. The noise sources are various and play a role at various frequency ranges. At lower frequencies, say below 1 kHz, noise sources are formed by human (shipping) activities and marine biology, albeit that the latter may be a subject of study as well. In the high frequency range, above 40 -- 50 kHz, the dominant noise source is thermal noise. Then, in the range of our interest, between 1 and 50 kHz, the dominant noise source is the so-called deep sea state noise. It has been found that the noise in the deep sea is strongly correlated with the wave height and the wind force in particular. This noise source has been parametrised by Knudsen \cite{Knudsen48}. Here we used the parametrisation as given by \cite{Kurahashi08} that is valid in the range between 1 and 40 kHz. The power spectral density ($P$) decreases with the frequency ($f$) and is given by:
\begin{equation}
\label{eqn:dss}
P (f, n_s) = 10 \log(f^{-5/3}) + 30 \log(n_s + 1) + 94.5 
\end{equation}
where, $n_s$ is a continuous variable that (when discretised) denotes the deep sea state. The lowest noise level is called deep sea-state zero (DSS0) corresponds to $n_s = 0$. 

Note that when a hydrophone system can be made with a sensitivity higher than DSS0, acoustic measurements become ocean noise limited. To reach this sensitivity is the main goal of the R\&D of hydrophones for cosmic ray particle detection. At 10 kHz the DSS0 has a power spectrum of 30 dB re $\mu$Pa/$\sqrt{{\rm Hz}}$.

\section{Fibre-based hydrophone system}
\label{sec:fibres_at_TNO}
\subsection{Fibre based hydrophones}
Fiber optic sensors are well known in literature as a component for optical acoustic sensors. In this the core component is a fibre laser sensor acting as transducer for the acoustic pressure. Using a Fibre Bragg Grating (FBG) inside an (Erbium) doped fibre  a laser system inside the core of the fiber is created. This technology is the so-called Distributed Feed Back fibre laser.
Changes of the cavity length (due to pressure) results in wavelength shift of the laser (sensor). 

\begin{figure}[ht]
 \begin{minipage}[]{0.6\textwidth}
   \begin{center}
   \includegraphics[width=0.95\textwidth]{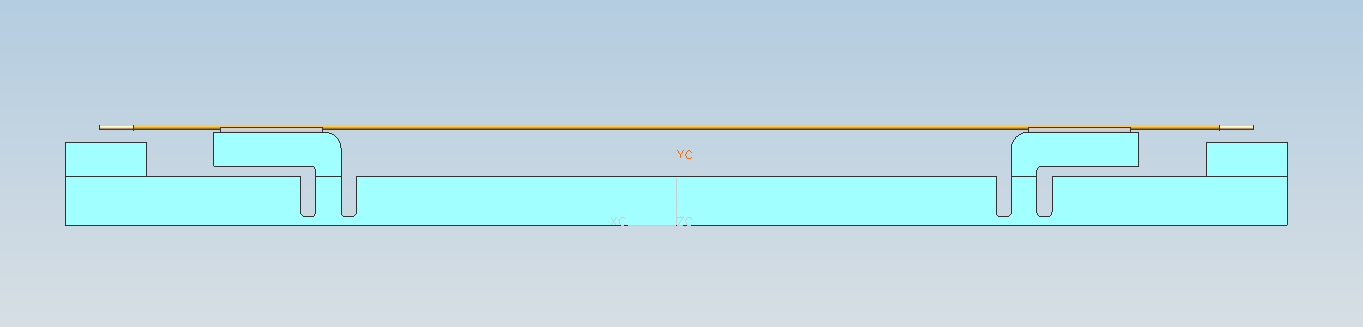}
   \end{center}
  \end{minipage}
  \begin{minipage}[]{0.4\textwidth}
   \begin{center}
     \includegraphics[width=0.95\textwidth]{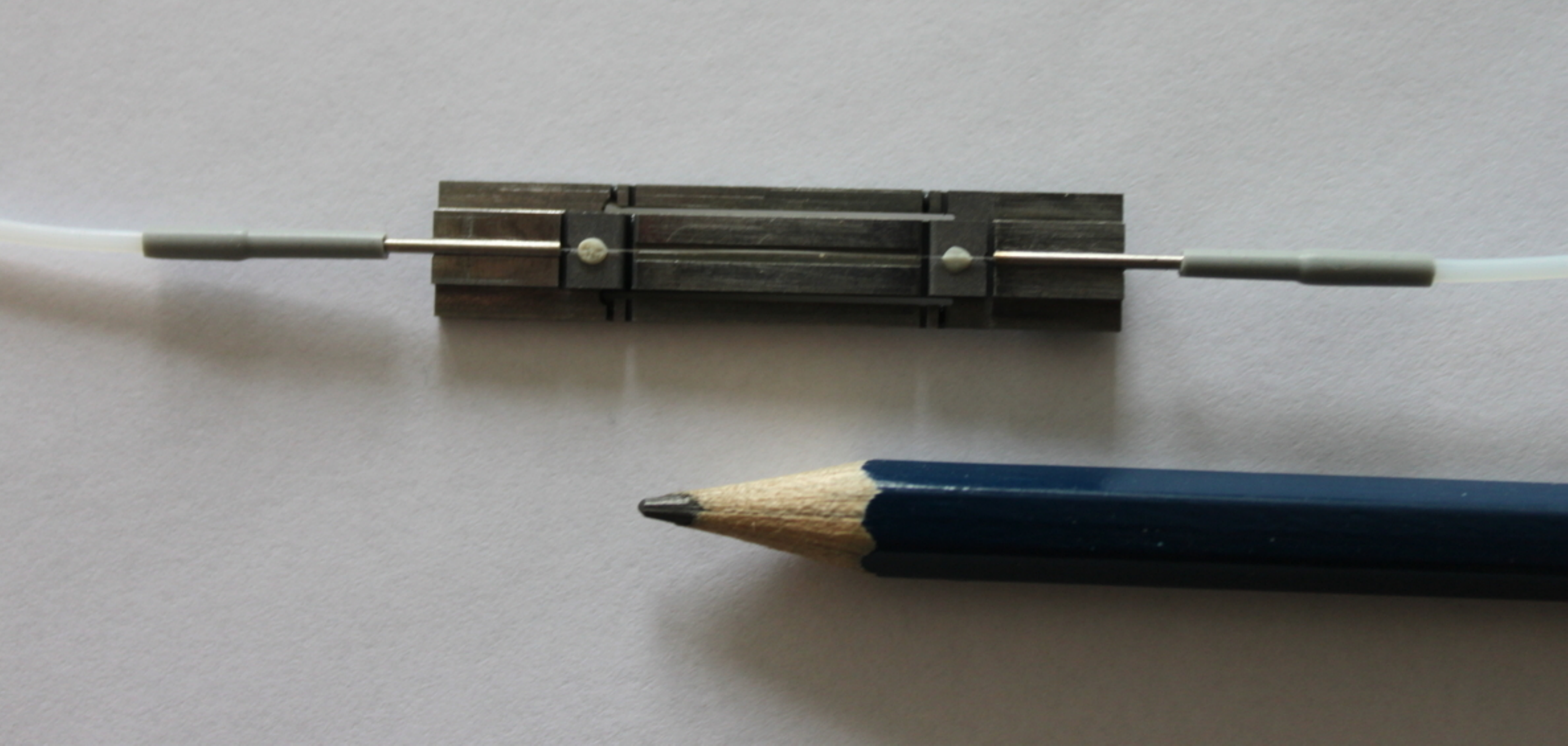}
   \end{center}
  \end{minipage}\\ \\
  \begin{minipage}[]{0.6\textwidth}
    \begin{center}
      (a)
    \end{center}
  \end{minipage}
  \hfill
  \begin{minipage}[]{0.4\textwidth}
    \begin{center}
      (b)
    \end{center}
  \end{minipage}
  \begin{center}
    \begin{minipage}[b]{\textwidth}
      \caption[]{\label{fig:hydrophone_TNO} Drawing and photograph of the hydrophone sensor as developed at TNO. Panel (a) shows the schematic drawing of the lamella of the sensor. The fibre that is glued on to the is depicted in yellow.  In (b) a photograph of the top view of the sensor is shown. Also the fibre and the two glue dots can be seen.}
  \end{minipage}
 \end{center}
\end{figure}

At the position of the FP cavity the fibre is glued on a sensor (or mechanical transducer). In principle, many cavities can inserted in a single fibre  so that up to 20 sensors can be read out simultaneously. The sensor as developed at TNO is shown in figure \ref{fig:hydrophone_TNO}. From the figure in the left panel it becomes clear that when an external pressure signal falls onto membrane of the transducer, it will induce strain on the cavity of the fibre. The photograph in the right panel of figure \ref{fig:hydrophone_TNO} shows the top view of the sensor. A pencil is shown next to indicate the size of the sensor.

\begin{figure}[ht]
  \begin{center}
    \includegraphics[width = 0.9\textwidth]{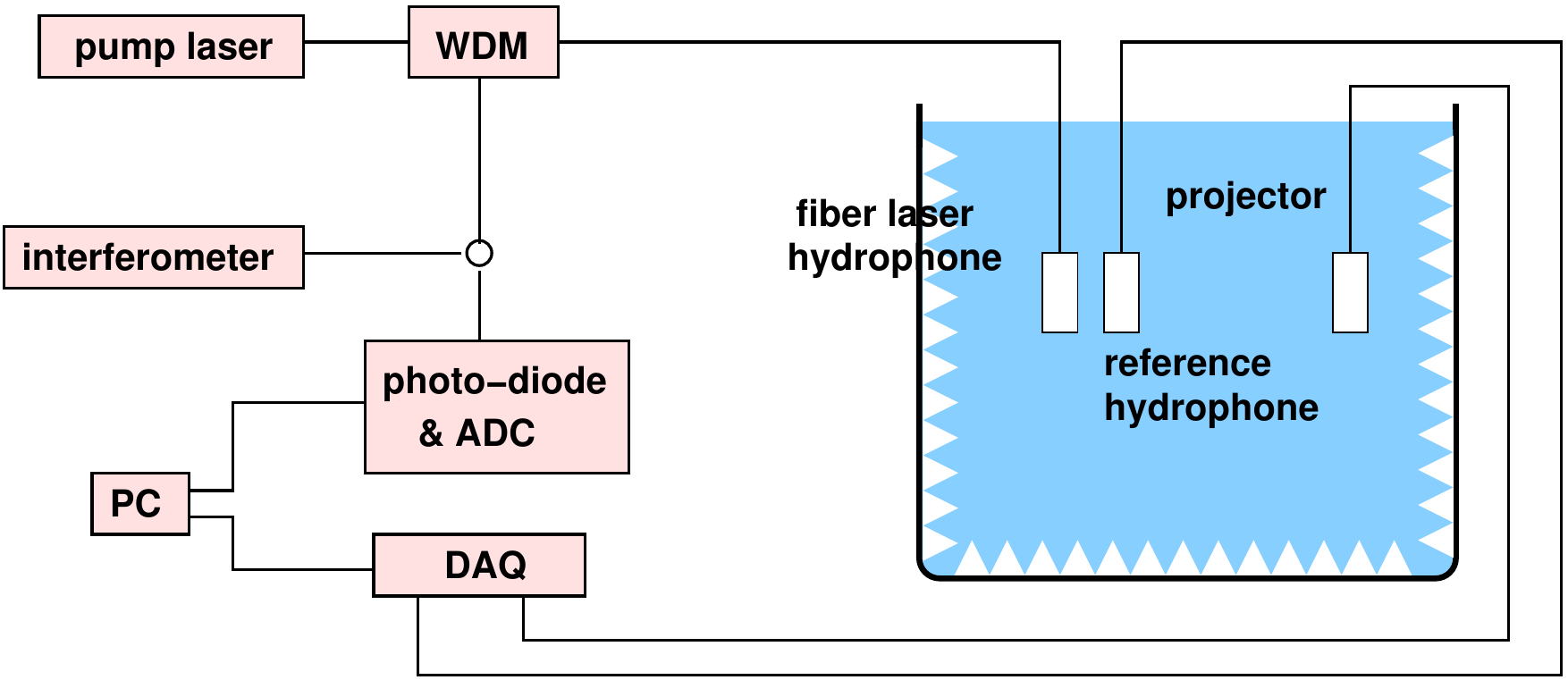}
    \caption[]{\label{fig:interogator} Schematic principle of an interrogator for the read out of a fibre laser hydrophone.}
  \end{center}
\end{figure}

The principle of interrogation is depicted in figure \ref{fig:interogator}. The wavelength shift that is induced by the strain in the cavity will be converted in to a phase shift using an interferometer. This interferometer is consists of a modified Mach-Zender interferometer that is designed to have a sensitivity (in combination with the sensor) of -154dB re 1 rad/$\mu$Pa, normalised to 1 meter optical path difference. Phase shift in the light output of the interferometer will result in light intensity variation in the photo-diodes. These diodes are housed on custom made PCB boards that also include preamplifiers. Finally the signal is digitized and multiplexed using a 20 bit ADC and stored on to a PC. The sampling rate that was used during the measurements was 50 kHz.

Light source is a 980 nm pump laser. The pump laser is a commercial available laser which is set to produce a power of about 200 mW. A narrow line width of about only 10 $\mu$W is then emitted by the fibre laser. 

\section{Hydrophones characterisation}
\label{sec:results}
\subsection{Experimental set up}
The hydrophones were characterised in an anechoic basin at TNO. This facility is dedicated to underwater acoustics and it mainly consists of a large acoustically isolated basin of dimensions $10\times8\times$8 m$^3$.  The experimental set-up is depicted in figure \ref{fig:interogator}. The fibre lasers were attached to a 50 m long standard optical fibre (type SMF28).

\begin{figure}[ht]
  \begin{center}
    \includegraphics[width = 0.6\textwidth]{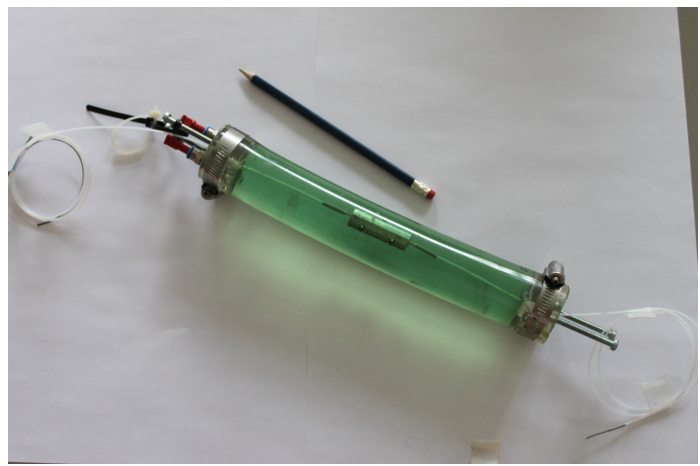}
    \caption[]{\label{fig:hose} Sensor located in an oil filled hose.}
  \end{center}
\end{figure}

Two identical fibre laser hydrophones were used in the experiments. One hydrophone was placed in the basin without any mechanical support or housing, while the second hydrophone was placed in a short oil-filled hose. The hose has a diameter of 40 mm and is 240 mm in length. The oil that was used to fill the hose, was Shell Naturel HF-E 15. It has a natural viscosity of 14.9 mm$^2$/s at 40$^0$C and a density of 0.892 kg/l at 20$^0$C. A photograph of the hydrophone as implemented in the hose is shown in figure \ref{fig:hose}. 
The photograph shows the straightforward implementation of a fibre based hydrophone in an experimental set-up that would consist of streamers.  

Two commercial, pre-calibrated piezo-based hydrophones, from Br\"uel\&Kjaer, type B\&K 8101 and  B\&K 8103 were used as a reference hydrophone. 
In addition, two acoustic sources (also called projectors) have been used during the measurements, which each there specific performances. The projector for the measurements at lower frequencies was an Actran projector (type LFPX-4). A B\&K 8105 hydrophone was used as a projector for the higher frequencies. These projectors and its driver software allowed to generate several acoustic spectra, such as frequency sweeps or pure tones.

\subsection{Results}
In this section the characterisation of the fibre based hydrophones is described.

\begin{figure}[ht]
  \begin{center}
    \includegraphics[width = 0.6\textwidth]{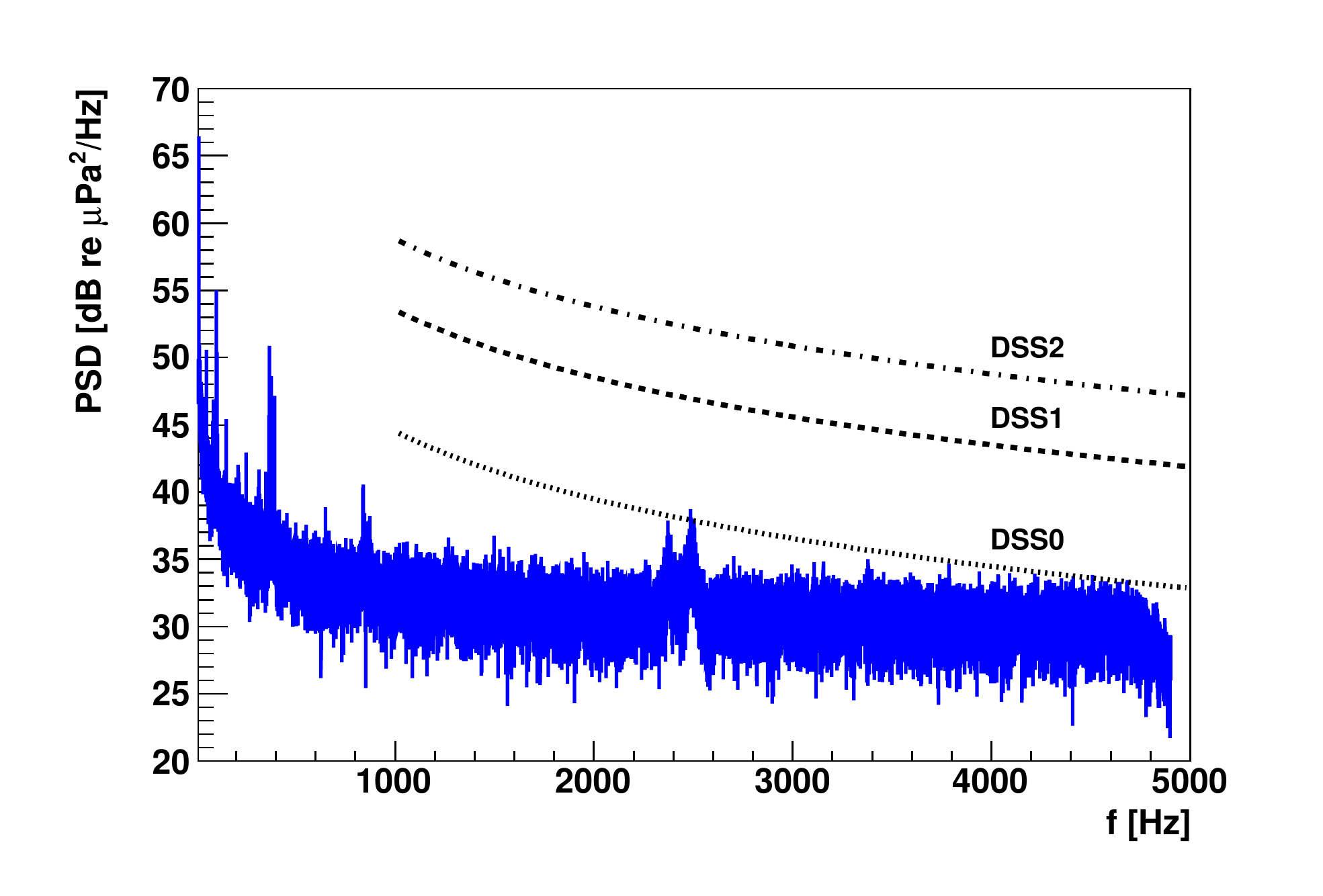}
    \caption[]{\label{fig:selfnoise} Self noise of the laser fibre based hydrophones.}
  \end{center}
\end{figure}

\paragraph{Sensitivity and linearity} $\;$\\
Prior to the characterization measurements of the hydrophone in the water basin, the self noise of the device was studied in a vibration isolated environment. Care was taken in the laboratory set-up to avoid as many noise sources as possible. In figure \ref{fig:selfnoise} we show the self noise of the laser fibre based hydrophone.

\begin{figure}[ht]
 \begin{minipage}[]{0.55\textwidth}
   \begin{center}
   \includegraphics[width=0.95\textwidth]{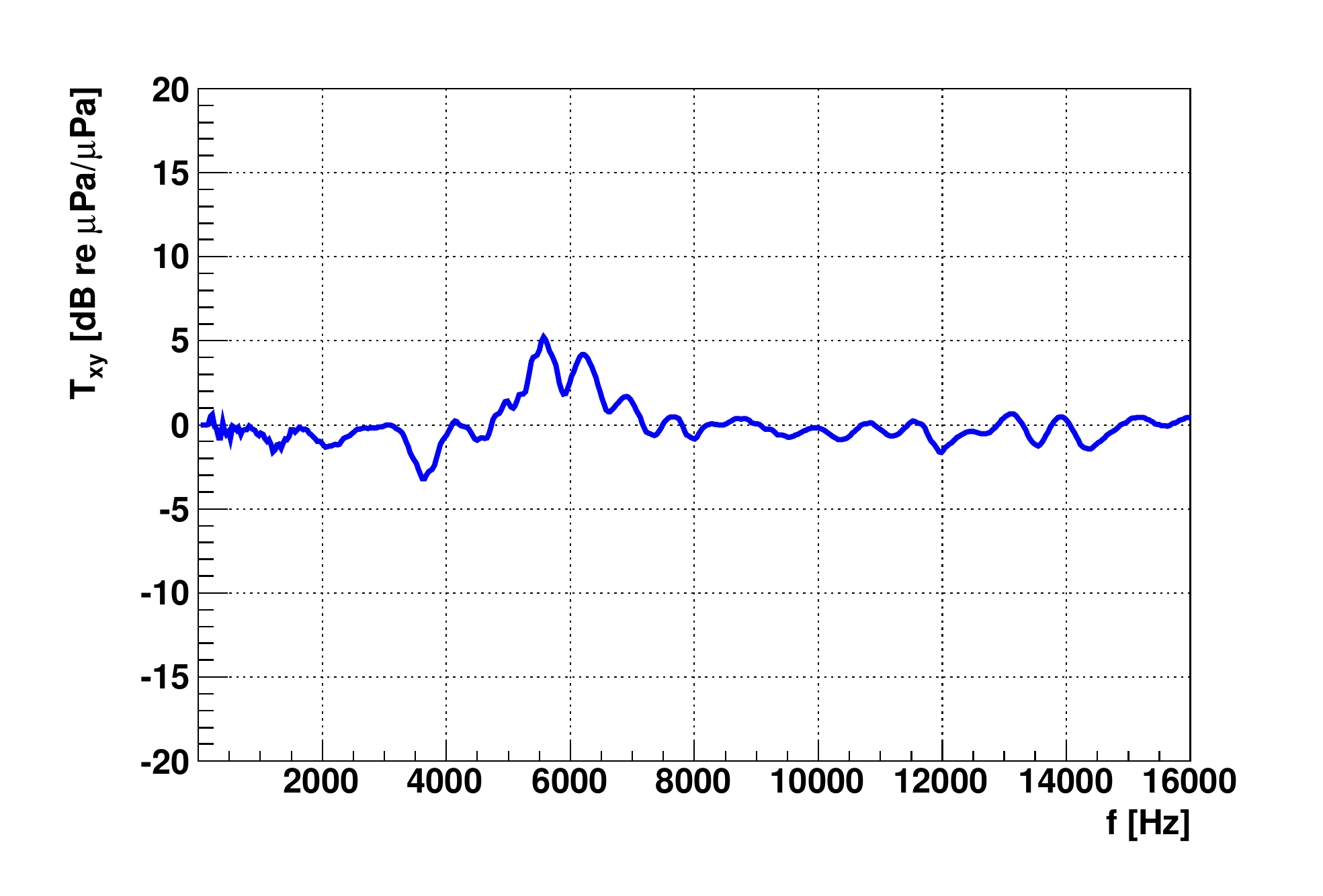}
   \end{center}
  \end{minipage}
  \begin{minipage}[]{0.55\textwidth}
   \begin{center}
     \includegraphics[width = 0.95\textwidth]{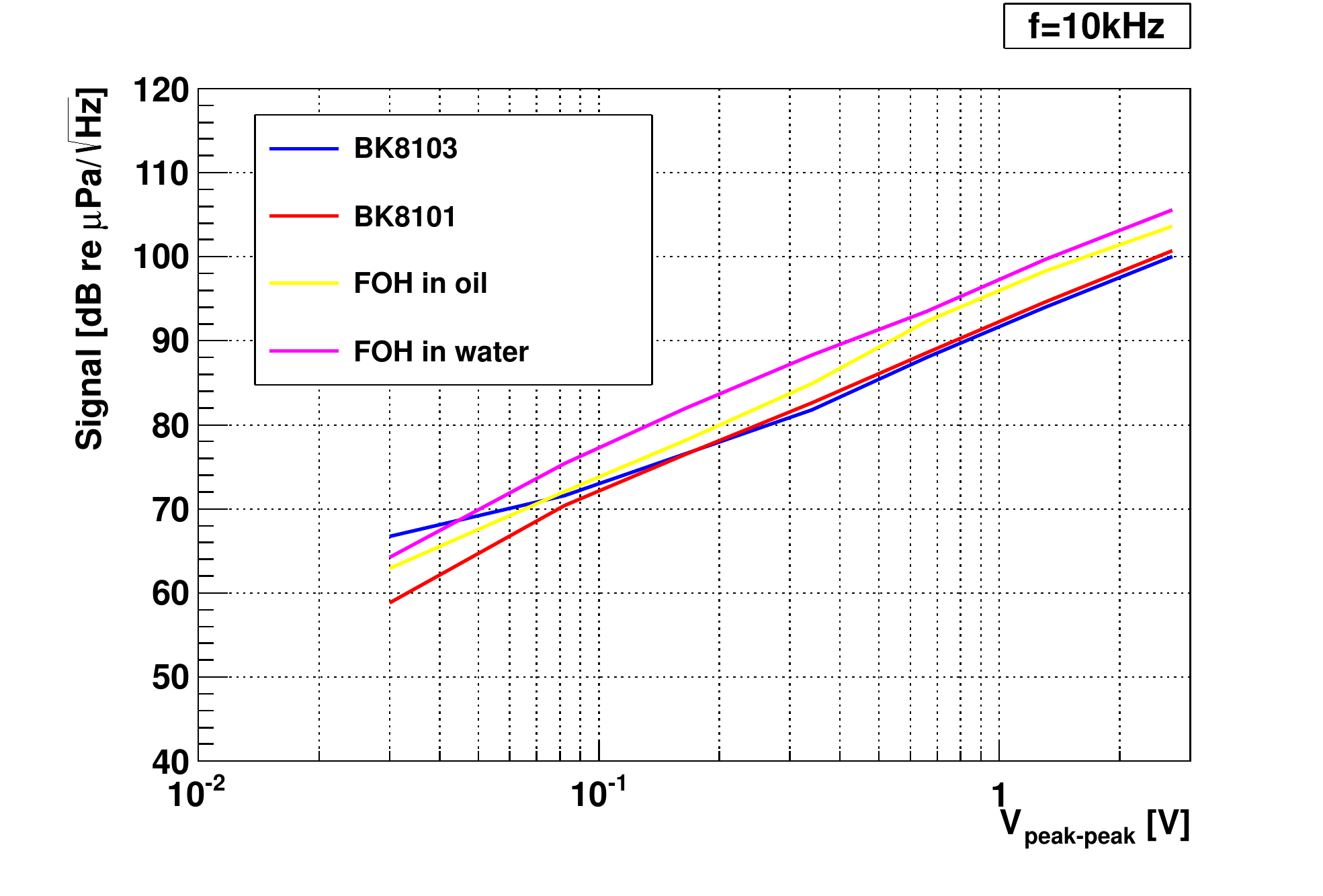}
   \end{center}
  \end{minipage}\\ \\
  \begin{minipage}[]{0.55\textwidth}
    \begin{center}
      (a)
    \end{center}
  \end{minipage}
  \hfill
  \begin{minipage}[]{0.55\textwidth}
    \begin{center}
      (b)
    \end{center}
  \end{minipage}
  \begin{center}
    \begin{minipage}[b]{\textwidth}
      \caption[]{\label{fig:linearity}(a) Response of the fibre optic hydrophone (in oil) compared to the reference hydrophone.  In (b) the linearity of several hydrophones as a function of the input current of the projector is plotted.}
    \end{minipage}
 \end{center}
\end{figure}

In figure \ref{fig:linearity}  the response of the fibre laser hydrophone is shown compared to the reference hydrophone. The plot shows a flat response curve in the range of interest. A small increase in the responsivity of the fiber laser was found around 6 kHz.
To determine the linearity of the hydrophone, a single tone was generated at a frequency of 10 kHz. The input to the projector was registered as being the input current to generate the sound and is shown in the right panel in figure \ref{fig:linearity}. The figure shows a linear behaviour of the hydrophones under test. All hydrophones are linear down to low levels to which the linearity of the projector was assured. One reference hydrophone (the B\&K 8103), however, showed some non-linearity was due to self-noise of the device. 

\paragraph{Pulse reconstruction}$\;$\\
To simulate the pulse that can be expected from a cosmic ray, a large number of bipolar pulses were generated and stored in a sound file. This sound file was played underwater. It was made sure that the pulse had the amplitude and duration according to the simulated pulses that have been recorded in literature \cite{Lahmann11}. The height of the pulse was verified using the  calibrated B\&K (type 8101) hydrophone.

\begin{figure}[ht]
  \begin{center}
    \includegraphics[width = 0.6\textwidth]{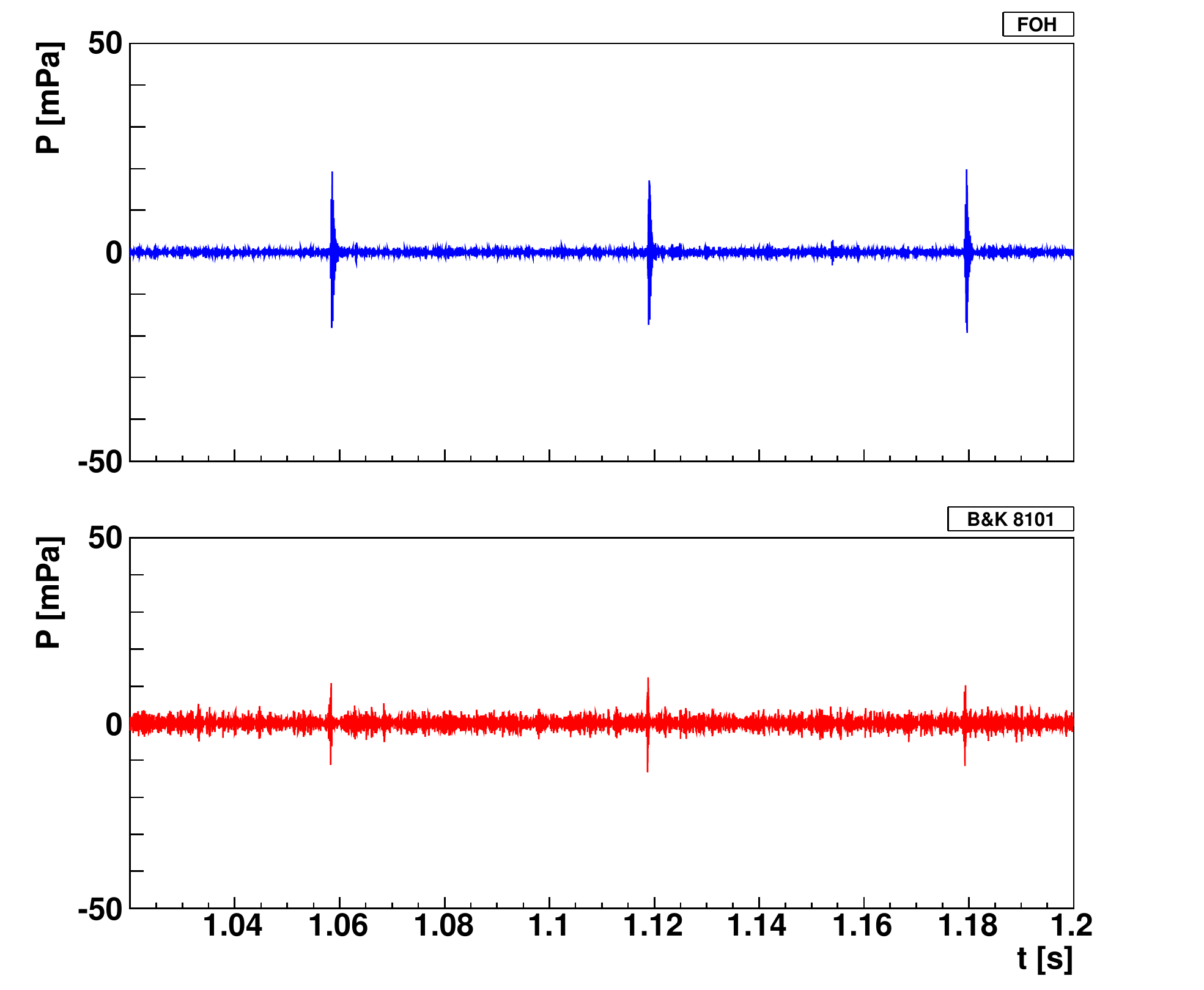}
    \caption[]{\label{fig:recorded_pulses} (Part of a) Time trace of both the fibre laser hydrophone (upper panel) as well as the B\&K hydrophone (lower panel).}
  \end{center}
\end{figure}

\begin{figure}[ht]
  \begin{center}
    \includegraphics[width = 0.6\textwidth]{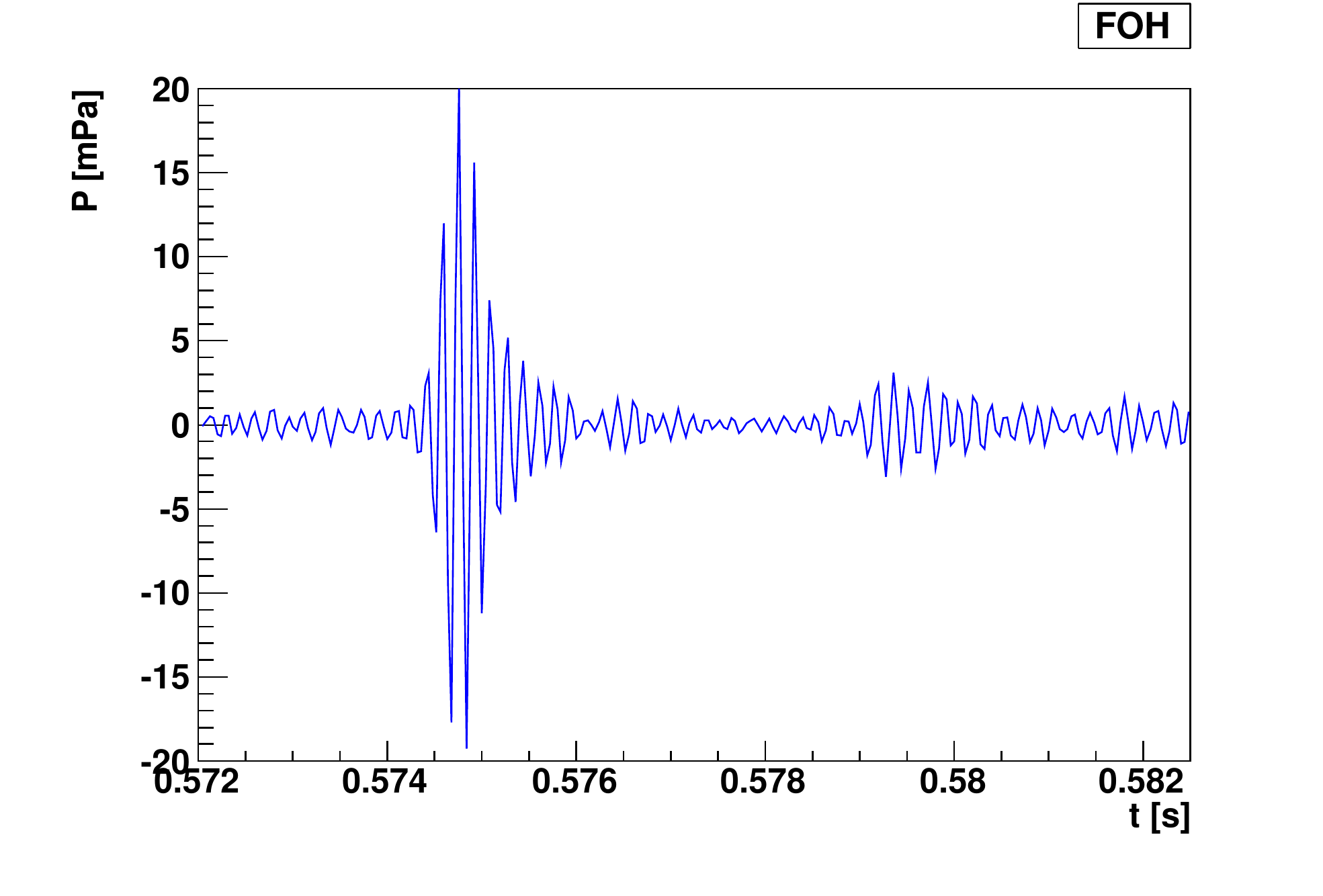}
    \caption[]{\label{fig:single_pulse} An example of a single pulse. The pulse has been recorded using the fibre laser hydrophone. Next to the recorded pulse also the echo can be seen with an amplitude of only several mPa.}
  \end{center}
\end{figure}

\begin{figure}[ht]
  \begin{center}
    \includegraphics[width = 0.6\textwidth]{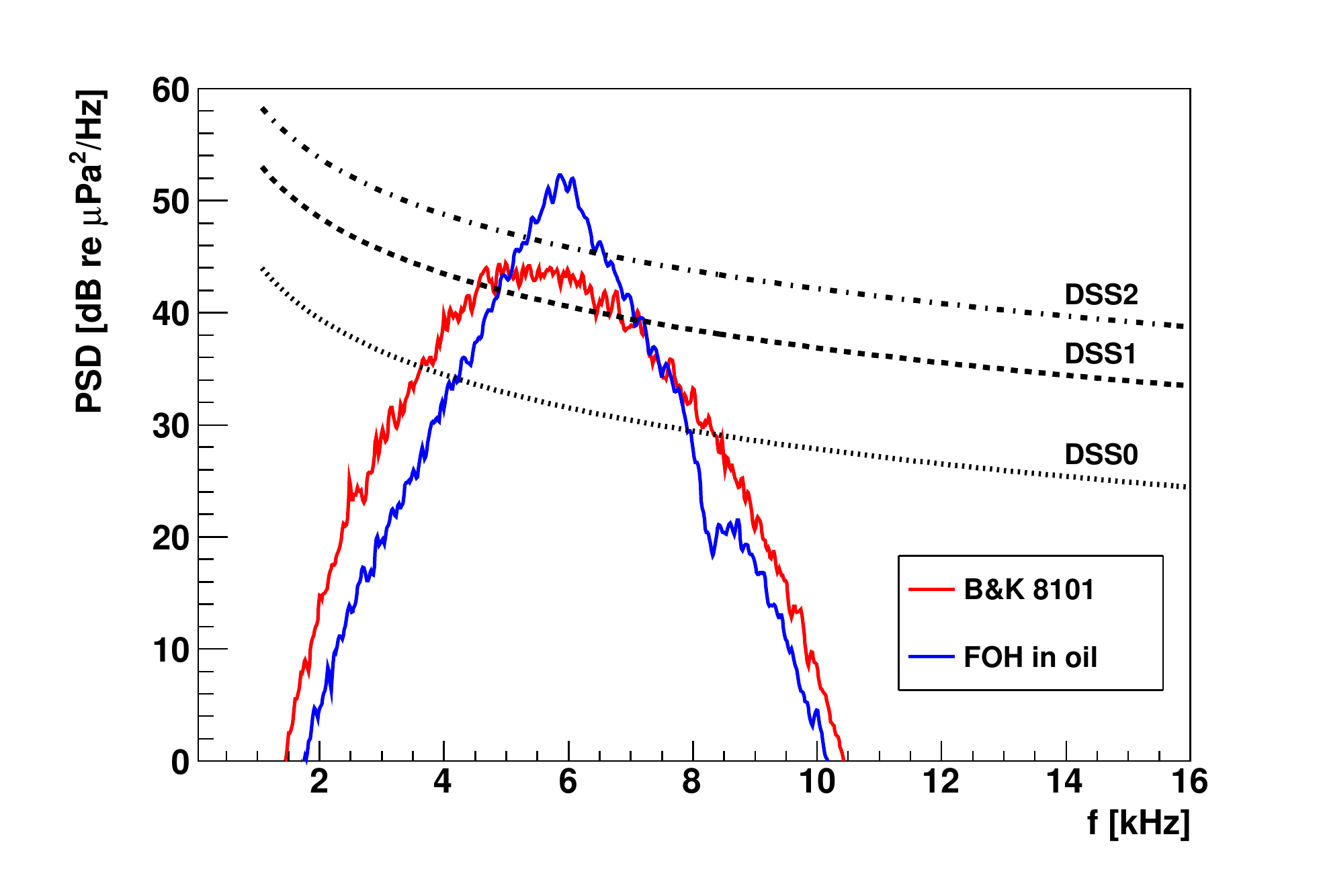}
    \caption[]{\label{fig:pulsespectrum} Power spectrum density of a time-trace of about 100 pulses. Also drawn are the noise levels, indicated as deep sea states (DSS).}
  \end{center}
\end{figure}

In figure \ref{fig:recorded_pulses} an example of a train of recorded pulses is shown the fibre laser hydrophone and the reference hydrophone. Here, only a simple pass-band filter between 5 and 7 kHz was applied (i.e.~4th order Butterworth filter) to pronounce the peaks. For both hydrophones the peaks clearly stand out of the background.
If we zoom in on a single pulse as shown in figure \ref{fig:single_pulse}, we see that the fibre laser hydrophone shows some ringing at the tail of the pulse. The origin of the ringing is yet unknown.
A power spectrum density was constructed from a time trace of about 100 pulses. The resulting plot is shown in figure \ref{fig:pulsespectrum}. As can be seen from the plot, the signal peaks at around 6 kHz which corresponds to the frequency of the pulse as shown in figure \ref{fig:single_pulse}. Furthermore the shape is dominated by the Butterworth filter. The spectrum of the fibre laser hydrophone shows a small dip around 8 kHz, which is due to the (finite) length of the oil hose in which the hydrophone was located.

\section{Conclusions}
\label{sec:conclusions}
Acoustic detection of cosmic rays in water is a promising means to study the high-end of the energy spectrum of cosmic neutrinos. The origin of the most energetic neutrinos is still a mystery and is a subject of investigation as well as a possible signature of the GZK cut-off limit in the neutrino spectrum. To detect the high energy cosmic neutrinos poses a challenge, due to the low expected flux and the low interaction cross-section of neutrinos. Therefore, the experimental set up that is used to study the cosmic neutrinos is required to have a large sensitive volume. One way to meet this requirement is to utilise the fact that particle showers induce a thermo-acoustic signal in water, which can be detected using hydrophones.  
 
A hydrophone detection system based of fibre laser optics has been designed and realised. The benefits of this technology are that the sensors are cost effective and so that a large volume can be equipped with a large number of sensors. Secondly, the sensor are straightforward to implement in long strings and require a minimum of front-end electronics. Finally and most important, the optical fiber technology allows to construct an extremely sensitive sensor.

To investigate the suitability of an application in a cosmic ray set-up, a measurement campaign was conducted to characterise the hydrophones. It has been found that fibre laser hydrophones exhibit a low self-noise level, which corresponds to level that is comparable to the lowest level of ambient noise in the ocean, the so-called deep sea state zero (DSS0). Hence a hydrophone system could be constructed that is ocean noise limited.

\end{document}